\documentclass[osajnl,twocolumn,showpacs,superscriptaddress,10pt]{revtex4-1} %% use 11pt for Applied Optics

\usepackage{amsmath,amssymb,graphicx}
 \usepackage[latin1]{inputenc}
 \usepackage{srcltx}

\begin{document}

%%%%%%%%%%%%%%%%%%%%%%% begin %%%%%%%%%%%%%%%%%%%%%%%%%%%%%%

\title{Speckle decorrelation in Ultrasound-modulated optical
tomography made by heterodyne holography}

\author{  M. Gross}

%\affiliation
\address{Laboratoire Charles Coulomb  - UMR 5221 CNRS-UM2
Universit\'e Montpellier
Place Eug\`ene Bataillon
34095 Montpellier, France }

%\date{\today}

\begin{abstract}
Ultrasound-modulated optical tomography (UOT) is a technique that images optical contrast deep
inside scattering media. Heterodyne holography  is a promising tool  able to detect the UOT tagged photons with high efficiency. In this work, we describe theoretically the detection of the tagged photon in heterodyne holography based UOT, show how to filter the untagged photon discuss, and discuss  the effect of speckle decorrelation. We  show that  optimal  detection sensitivity can obtain, if the frame exposure time is of the order of the decorrelation time.
\end{abstract}

%\ocis{170.1650, 170.3660, 290.7050, 090.0090, 170.7050}

\maketitle
OCIS codes: 170.1650, 170.3660, 290.7050, 090.0090, 170.7050

%\section{Introduction}

Light scattering prevents optical imaging from achieving high resolution inside
scattering media deeper than about 1 mm in tissue.  Ultrasound-modulated
optical tomography (UOT)\cite{wang1997ultrasound,elson2011ultrasound} also called acousto-optic imaging \cite{resink2012state}, has been developed to overcome this limit by combining ultrasonically defined spatial resolution and optical contrast (i.e. sensitivity to the bulk optical properties like absorption).  One of the purpose of the technique is to use the optical contrast to detect breast tumors that cannot be seen with ultrasound, because the ultrasonic contrast is too low.
In an UOT experiment, the light scattered through a diffusing sample crosses an
ultrasonic beam, and, due to the acousto-optic effect,  undergoes a frequency shift equal to the ultrasonic frequency \cite{leutz1995ultrasonic,wang2001mechanisms}. % By
%detecting the frequency-shifted light (tagged light), ultrasonically defined spatial resolution can
%be reached.
 By detecting the  frequency-shifted  photons, called tagged photons, and by plotting their weight  as a function of the ultrasonic  beam geometry, 2D or 3D images of the sample can then be obtained with ultrasonic spatial resolution.

Various methods have been developed to detect the very low tagged
photons signal out of a large background of untagged photons \cite{elson2011ultrasound,resink2012state}. First experiments use  single-pixel detector   and detection of the tagged photon AC modulation at the ultrasonic frequency \cite{wang1995continuous,leutz1995ultrasonic,kempe1997acousto}.  Since each speckle grain oscillates with a different phase, the single pixel method detects, with a good efficiency, no more than one speckle grain. This
severely limits the detection etendue (defined as the product of the detection area and the
acceptance solid angle).
To increase the detection etendue without reducing the modulation
depth, three types of methods have been developed. The first type relies on incoherent detection with a narrow spectral
filter ($\sim$MHz) that  filter out the untagged light. A large-area single-pixel detector can be
used. Examples include Fabry-Perot interferometers \cite{sakadvzic2004high,kothapalli2008ultrasound,rousseau2009ultrasound}
and spectral-hole burning \cite{li2008detection,li2008pulsed,zhang2012slow}
based
methods. These techniques %are immune to speckle decorrelation due to motions of scatterers,
require bulky and expensive equipment.
The second and third types of method use   interferences and  are thus  sensitive to  the  signal phase decorrelation  due to the living tissue  inner motions, and to the corresponding Doppler broadening. For breast, this broadening is $1.5$ kHz \cite{gross2005heterodyne}.
The second method is  based  on a photorefractive crystal,  which records the volume hologram of the sample  scattered field. This hologram can be then used to generate a diffracted  field able to interfere with  the  scattered field on a large area single-pixel detector \cite{murray2004detection,ramaz2004photorefractive,gross2005theoretical,lai2012ultrasound}. The method has a large optical etendue ($\sim 10^8$ speckle), but is somewhat sensitive to decorrelation, since the response time of the crystal is usually much longer than the speckle correlation. Promising results are expected with Sn2P2S6:Te and Nd:YVO4 crystals, because of their short response times \cite{farahi2010photorefractive,jayet2014fast}.

The third type of method uses a pixel array, i.e., a camera, to detect the UOT tagged photons \cite{leveque1999ultrasonic,yao2000frequency,li2002methods,li2002ultrasound}. The optical etendue ($\sim 10^5$ to $10^6$ speckle) is then related to the  number of pixels of the camera. The camera method has been improved by adapting the heterodyne holography technique \cite{le2000numerical} to the tagged photon detection  \cite{gross2003shot}. By tuning the  LO beam frequency near the ultrasonic sideband, and by using a properly adjusted spatial filter,  the tagged photons were  detected selectively.
Moreover, optimal noise detection was obtained,   since shot noise is the dominant noise in heterodyne holography \cite{gross2007digital,verpillat2010digital,lesaffre2013noise}.
%
%Heterodyne holography UOT is can be thus considered as an ideal detection scheme of the tagged photons.
%
The reference \cite{gross2003shot} experiment    was nevertheless performed with a phantom sample, whose decorrelation is low, and it is generally considered that the  heterodyne holography UOT method cannot be used with a living sample,  whose speckle decorrelation time $\tau_c$   is shorter than the time needed to record four camera frames (where $\tau_c= 0.1$ ms for the light scattered "in vivo" through a woman's breast \cite{gross2005heterodyne}).  Resink et al. \cite{resink2012state} wrote, for example,  "all frames of the one to four phases [i.e. the four frames of the camera] should be taken within the speckle decorrelation time".

\begin{figure}[h]
\begin{center}
  \includegraphics[width=6.5cm]{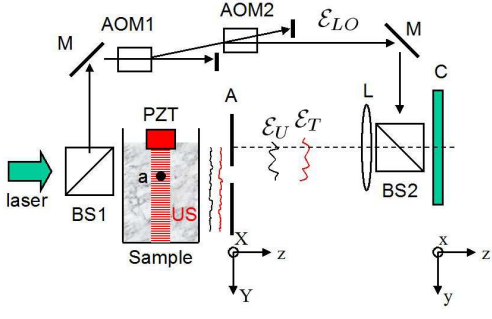}
  \caption{Typical UOT setup: BS1, BS2: beam splitter;  M: mirror; AOM1,AOM2: acousto optic modulator; PZT: piezoelectric transducer that generates the ultrasonic beam US; a: absorber embedded in the diffusing  sample; A: rectangular aperture; L: lens of focal $d$; C: camera; ${\cal E}_{LO}, {\cal E}_{T},{\cal E}_{U}$: LO, tagged and untagged fields. }\label{Fig_setup}
\end{center}
\end{figure}

In this work, we  analyzed theoretically the   heterodyne holography  UOT detection scheme,
and we calculated how   untagged photons, speckle noise,   shot noise, decorrelation and  etendue,  affect the UOT signal. By adjusting the calculation parameter, we got   results very similar to the ones of \cite{gross2003shot}. By comparing results obtained with and without decorrelation, we showed that the Resink et al. remark is not valid, and  that heterodyne holography  remains,   with  decorrelation, an optimal detection scheme  of the tagged photons. Note that this point was already demonstrated     for  the detection of the untagged photon in experiments done without ultrasound \cite{atlan2006frequency,atlan2006laser,lesaffre2006effect,atlan2007cortical,atlan2007spatiotemporal,atlan2008high}.

To introduce our theoretical discussion,  let us consider a typical heterodyne
 holographic UOT  setup (Fig. \ref{Fig_setup}). %\cite{gross2003shot}.
A laser of frequency $\omega_L$ is split  by the  beam splitter BS1  into a signal beam and a local oscillator (LO) beam.
The signal beam  travels trough the diffusing sample S and is scattered by it.
The sample is explored  by an ultrasonic beam US of  frequency $\omega_{US}$.
 The light transmitted by the sample exhibit to components. The first component  at $\omega_T=\omega_L + \omega_{US}$ is weak ($\sim 10^{-2}$ to $10^{-4} $ in power), and  corresponds to the tagged  photons  that have interacted with the ultrasonic (US) beam.   The second  component  at  $\omega_U=\omega_L $ is the main one ($\simeq 100 \%$ in power). It  corresponds to untagged photons which  have not interacted with  US.
%
%To detect selectively the tagged photons, and to perform this detection with optimal sensitivity,

A  rectangular aperture A, located off axis near the sample, control the size  and location of the zone of the sample where   of the tagged and untagged fields ${\cal E}_T$ and ${\cal E}_T$  are detected.
The  ${\cal E}_T$ and ${\cal E}_T$ fields are mixed with the LO field ${\cal E}_{LO}$ by the beam splitter BS2, and the camera C records a sequence of $M$  frames $I_m$ (with $m=0...M-1$) corresponding to the   interference pattern:  ${\cal E}_T$ +  ${\cal E}_U$ + ${\cal E}_{LO}$,
$I_m$ being  recorded at time  $t_m = m \Delta t$, where $\Delta t=2\pi m/\omega_C$ is the pitch in time, and  $\omega_C$ is the camera frame frequency.
The hologram $H_C$ of the aperture A (that is back illuminated by ${\cal E}_T$ and  ${\cal E}_U$)  is calculated, in the camera plane C, by combining    frames $I_m$. The hologram $H_A$, in the aperture plane A, is then calculated from $H_C$. The signal of interest (tagged or untagged photon) is calculated from $H_A$.

To analyze theoretically the Fig.\ref{Fig_setup} experiment, let us define, in plane A and C,  the untagged, tagged and LO fields and their respective complex amplitudes, which are slow varying with time $t$:
\begin{eqnarray}
  {\cal E}_{A,U}(X,Y,t) &=&   E_{A,U}(X,Y,t)~e^{j \omega_L t} + \textrm{c.c. }\\
\nonumber   {\cal E}_{A,T}(X,Y,t) &=&   E_{A,T}(X,Y,t)~e^{j (\omega_L+\omega_{US}) t}+ \textrm{c.c. } \\
\nonumber   {\cal E}_{C,U}(x,y,t) &=&   E_{C,U}(x,y,t)~e^{j \omega_L t} + \textrm{c.c. }\\
\nonumber  {\cal E}_{C,T}(x,y,t) &=&   E_{C,T}(x,y,t)~e^{j (\omega_L+\omega_{US}) t} + \textrm{c.c. }\\
\nonumber  {\cal E}_{C,LO}(x,y,t) &=&   E_{LO}~e^{j \omega_{LO} t} + \textrm{c.c. }
\end{eqnarray}
where c.c. is the complex conjugate. Here, $X,Y$ are the  coordinates  in plane A, and $x,y$  in plane C.  To simplify theory, we have considered here that  $ E_{LO}$ do not depend on $x,y$ and $t$.
In plane A, the  tagged and untagged photon fields are  fully  developed  speckle. The complex fields  $ E_{A,T/U}(X,Y,t_m)$
are thus random Gaussian complex quantities   uncorrelated from one pixel $X,Y$ to any other $X',Y'$.
The random amplitudes $ E_{A,T/U}(X,Y,t_m)$
do not depend on $t_m$  without decorrelation, and are  uncorrelated from one frame (i.e. $m$) to the next (i.e. $m+1$) with decorrelation.

A lens $L$, which is located near the camera, and whose focal plane is close to plane A,   collects the  fields. Because of L, the tagged and untagged fields in planes C and A are related by a Fourier transform %
\begin{eqnarray}\label{Eq_propag_A2C}
 \nonumber
  E_{A,U/T}(X,Y) &=& {\tilde E_{C,U/T}}(k_x,k_y)=  \textrm{FFT} (E_{C,U/T}(x,y))
\end{eqnarray}
where  $(X,Y)= (k_x,k_y) \times |\textrm{CA}|/k$ with $k=2 \pi/ \lambda$. To simplify calculations, the discrete Fourier transform  (FFT) is made  within a  calculation grid that fits with the camera pixels in plane C. The pitch $\Delta x$ of the discrete coordinates $x,y$ is thus equal to the size of the pixel of the camera. Because of the FFT, the   pitch $\Delta X$ in plane A is
\begin{eqnarray}
 %\Delta k &=& 2 \pi /( N \Delta x  )  \\
  % \nonumber \Delta X&=& |\textrm{CA}| ~(\Delta k   /k)\\
    \Delta X&=& 2 \pi |\textrm{CA}| ~  /( N k \Delta x  )
\end{eqnarray}
where $N$ is the number of pixels of the camera ($N=1024$ typically).
The detection etendue is thus $G= S_A  S_D/ |\textrm{CA}|^2=N^2\lambda^2$, where $S_A=|N\Delta X|^2$ and $S_C=|N\Delta y|^2$  are the areas of the calculation grid in plane A and C. The number of modes or speckle grains that can be detected is thus equal to  the number of pixels of the camera  $N^2$.

The frame signal $I_m$ correspond to the sum of the tagged,  untagged and  LO photons.
To detect the tagged photons, $\omega_{LO}$  is made close to the tagged photon frequency $\omega_L+\omega_{US}$. The  LO and the tagged photons thus interfere (and are summed in fields), while the untagged photons do not interfere (and  are summed in intensities). We have thus:
  \begin{eqnarray}\label{Eq_Im}
        I_{m}(x,y) &=&  \left|E_{C,T}(x,y,t_m ) + c^m  E_{LO} \right|^2  \\
        \nonumber &&   ~~~~~+ \left|E_{C,U}(x,y,t_m )\right|^2
  \end{eqnarray}
where $c=e^{j(\omega_{LO}-\omega_{US}-\omega_{L})\Delta t}$ is the LO versus tagged photon shift of phase.
On the other hand,    to detect the untagged photons, $\omega_{LO}$  is  close to   $\omega_{L}$, and
$I_m$ is given by an  equation similar to Eq. \ref{Eq_Im},
where the indexes $U$ and $T$ are exchanged, and where  $c=e^{j(\omega_{LO}-\omega_{L})\Delta t}$.
Because of the random nature of light emission and camera photo conversion, the frame signal $I_m$ is affected by  shot noise yielding  $I'_m$:
\begin{eqnarray}\label{Eq_I'_m}
 I'_m(x,y)&=&  I_{m}(x,y) + s(x,y,m)\sqrt{I_{m}(x,y)} ~~)
\end{eqnarray}
where the term $s\sqrt{I_{m} }$ accounts for shot noise. Here,  $I_m$ must be expressed   in photo electron Units per pixel and per frame, while $s$ is a real Gaussian random variable of variance $ \langle s^2\rangle=1$ uncorrelated with pixels (i.e. $X,Y$)  and with frames (i.e. $m$).

The hologram $H_C$ is then calculated from  $I'_m$.
Without decorrelation, four phase detection of the tagged photons is made. We have thus:  $H_C=\sum_{m=0}^{M} j^m I_m$ (where $M$ frames is a multiple of four) and  $ \omega_{LO}= \omega_L+ \omega_{US}+ \omega_C/4 $ yielding $c=j$ in Eq.~\ref{Eq_Im}.
With decorrelation, two phase detection with two frames is made: $H_C=I_0-I_1$, and  $\omega_{LO}=   \omega_L+ \omega_{US} + \omega_C/2$ yielding $c=-1$.
In $H_C$, the holographic term of interest:  $E_{C,T} E^*_{LO}$ (where $^*$ is the complex conjugate operator),  is proportional to the field $E_{C,T}$. Because of L, the holograms  $H_A$  and $H_C$   in  planes A and C  are related by a Fourier transform %
\begin{eqnarray}%\label{Eq_H_A}
H_{A}(X,Y) = {\tilde H_C}(k_x,k_y)=   \textrm{FFT}~ (~H_{C}(x,y)~)
\end{eqnarray}
\begin{figure}
\begin{center}
  \includegraphics[width=3.8cm]{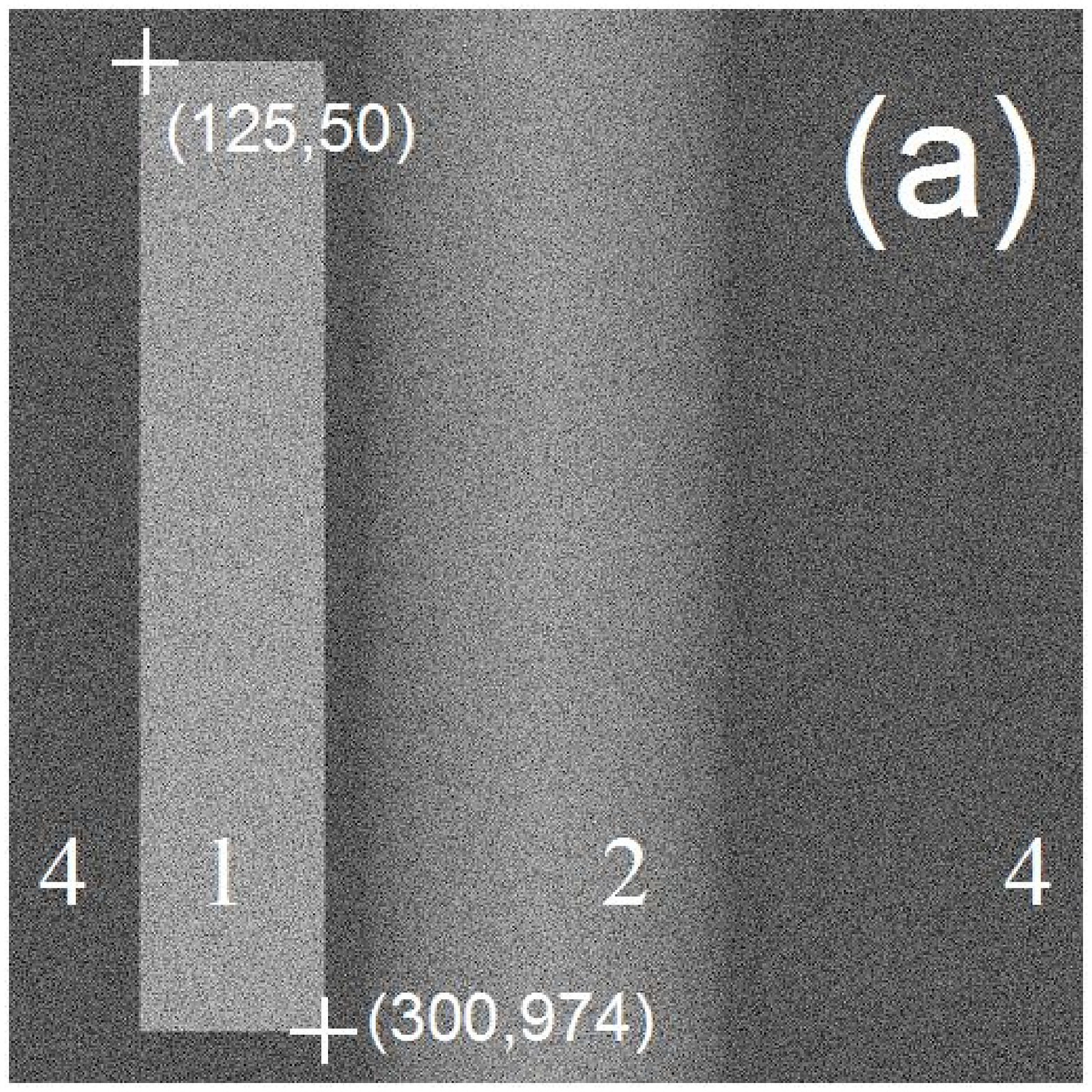}
 \includegraphics[width=3.8cm]{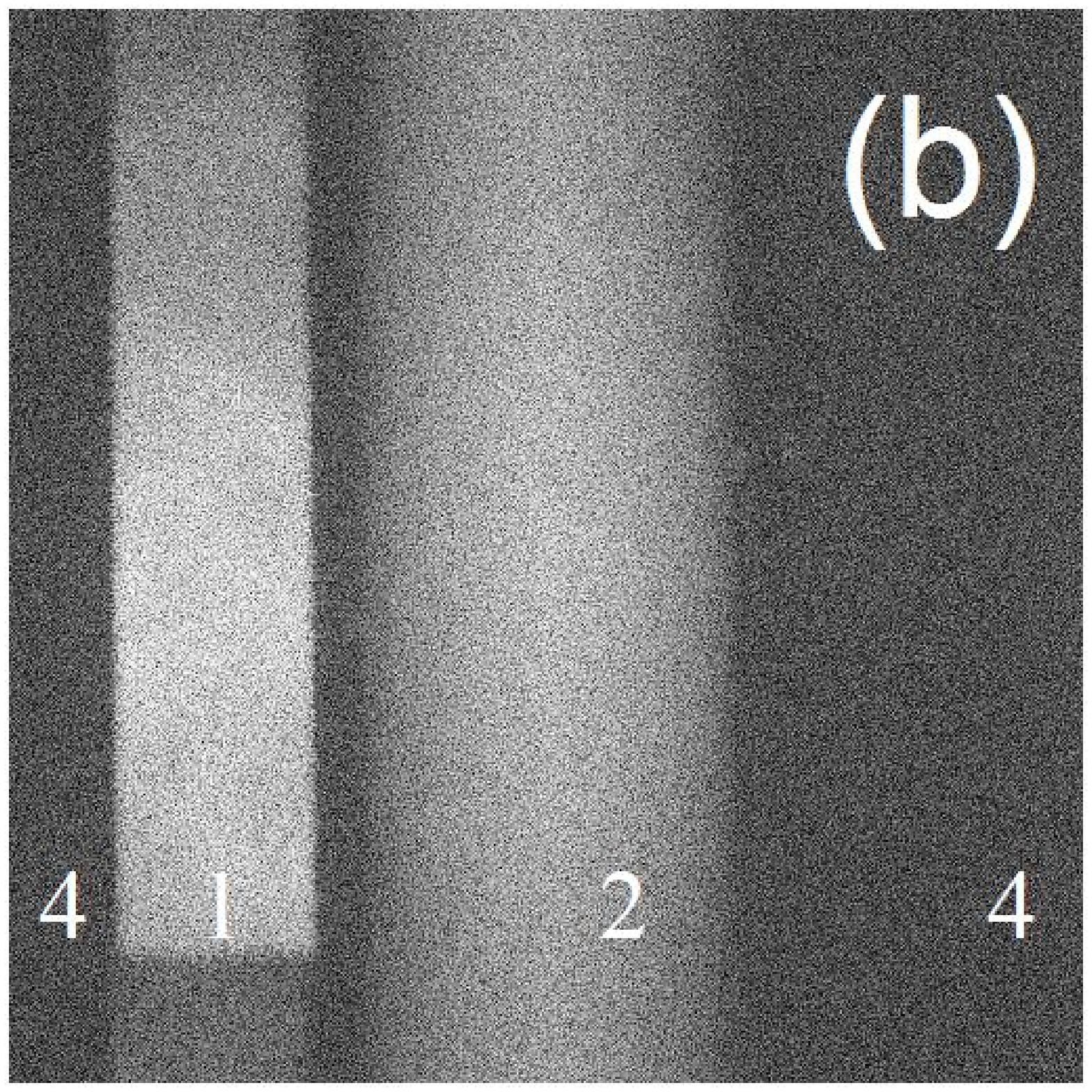}\\
   \includegraphics[width=3.8cm]{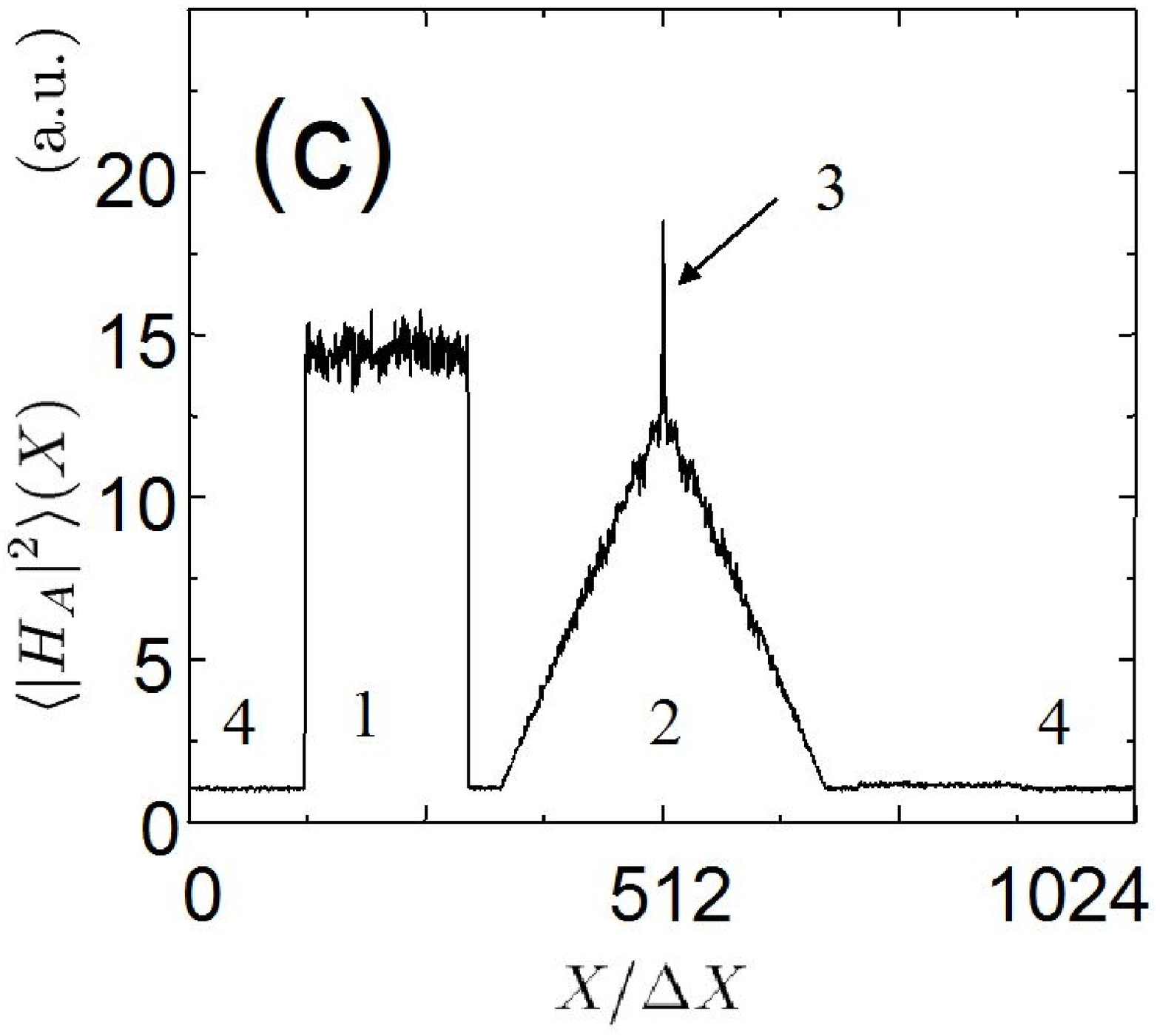}
  \includegraphics[width=3.8cm]{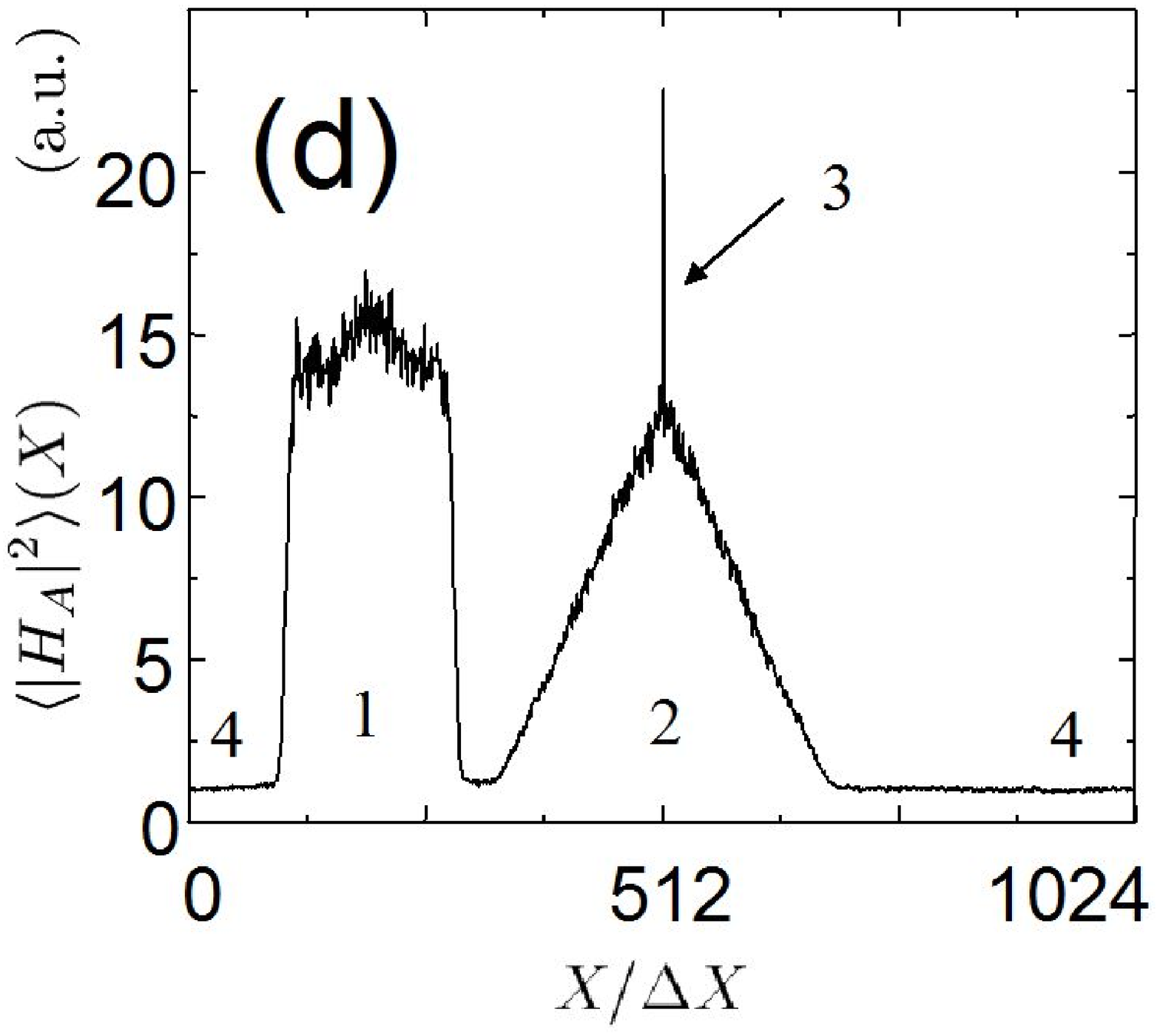}
  \caption{Tagged intensity images $|H_A(X,Y)|^2$ (a,b) and curves $\langle  |H_A(X)|^2 \rangle$  (c,d)  obtained by calculation (a,c) and from  ref.\cite{gross2003shot} experiment (b,d). The images $|H_A(X,Y)|^2$ (a,b) are displayed in an arbitrary logarithmic scale. The curves $\langle  |H_A(X)|^2 \rangle$  (c,d) are normalized with respect to the background. }\label{fig2}
\end{center}
\end{figure}
We have calculated, without decorrelation, the tagged photon hologram $H_{A}$ for  $M=12$ frames (like in \cite{gross2003shot}). The calculation is made with  $|E_{LO}|^2=10^4$,  $\langle| E_{A,T}|^2 \rangle =1.33$  and  $\langle| E_{A,U}|^2 \rangle =3 \times 10^3$  photo electron per frame and per pixel, where $\langle ~\rangle$ is the average over X and Y within the aperture.  Note that tagged and untagged  energies ($\sum_{\textrm{pixels}} | E|^2$) are    the same in planes A and C, because Eq.\ref{Eq_propag_A2C} conserves energy.  The  coordinates  of the upper left and bottom right aperture corners were $(125, 50)$ and $(300, 974)$.

We have displayed in Fig. \ref{fig2}  the arbitrary logarithmic scale intensity image $|H_A(X,Y)|^2$ obtained by   calculation~(a), and in experiment~(b)~\cite{gross2003shot}. Note that the calculation parameters were chosen here to fit with \cite{gross2003shot}.   To further compare our calculation with~\cite{gross2003shot}, we have calculated the curves $\langle |H_A(X)|^2 \rangle$:
\begin{eqnarray}\label{Eq_profile_HA}
  \langle |H_A(X)|^2 \rangle  &=& (1/N)  \sum_X |H_A(X,Y)|^2
\end{eqnarray}
Figure \ref{fig2} (c,d) show  the curves $\langle  |H_A(X)|^2 \rangle$ obtained by calculation~(c) and from~\cite{gross2003shot}~(d).  The curves are normalized with respect to the background that is obtained without tagged and untagged photons and  that corresponds to shot noise.
The good  agreement  with~\cite{gross2003shot} validates our theoretical calculation.

In figure \ref{fig2}, the tagged photon signal corresponds to the image of the aperture, i.e. to the bright rectangular zone 1, which is located in the left of  images (a,b), because the aperture is located of axis. The aperture corresponds also  to the rectangular walls  1, in  curves (c,d). On the other hand, the blurred bright zone 2, in the center of  (a,b), and the triangular wall 2, in  (c,d), corresponds to a parasitic detection of the untagged photon signal, which does not cancel here because of decorrelation (in experiment) and shot noise (in experiment and calculation). The parasitic detection of the LO fields yields a very narrow peak located in the center of the calculation grid, which is only visible on the curves (arrow 3). To the end, shot noise yields a flat background in all points of the images and the curve (zones 4).

\begin{figure}
  \begin{center}
   %Requires \usepackage{graphicx}
  %\includegraphics[width=4cm]{images/fig_ima_incorr_a}
%     \includegraphics[width=4cm]{images/Fig_recons_c}
%
  \includegraphics[width=4cm]{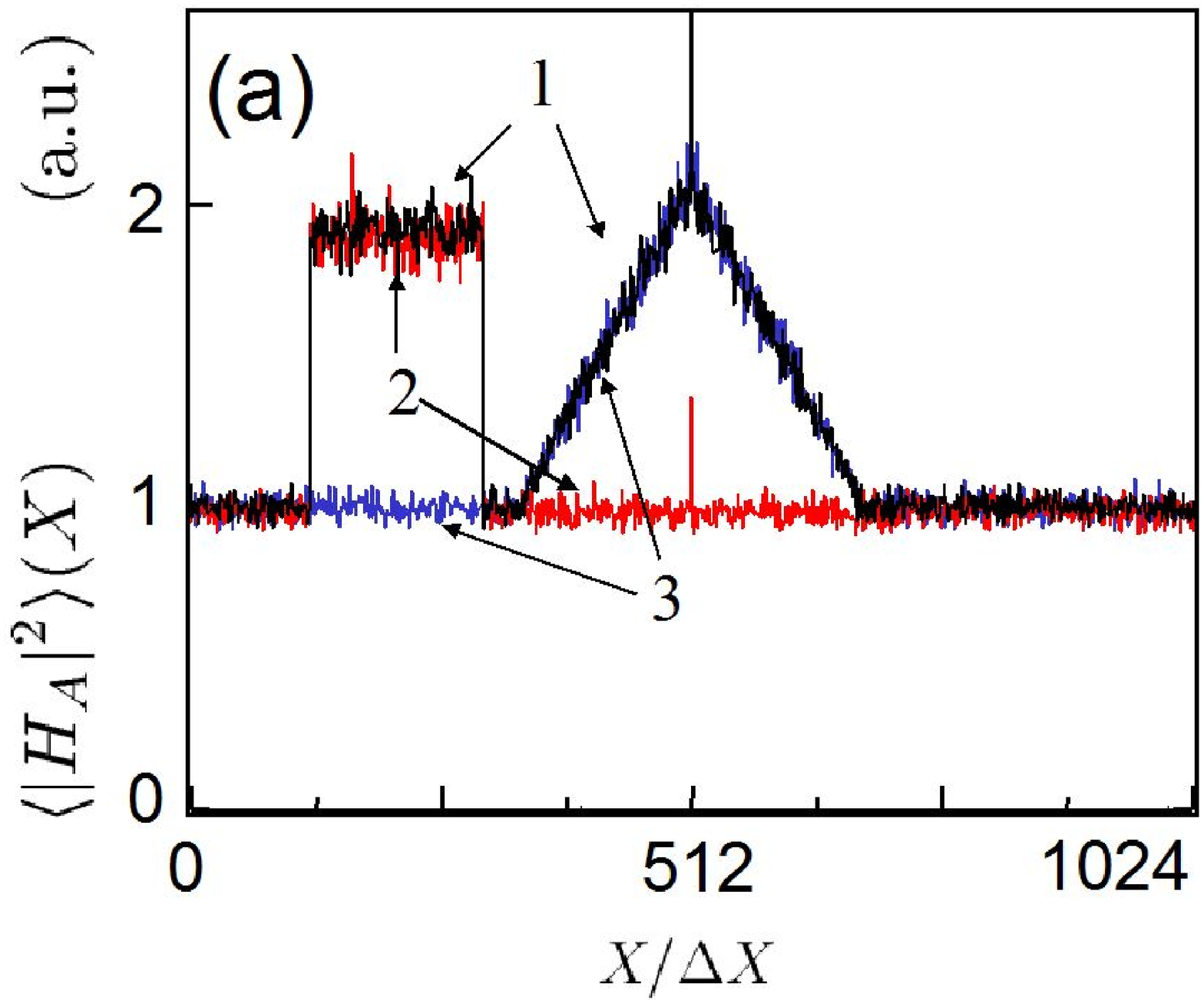}
   \includegraphics[width=4cm]{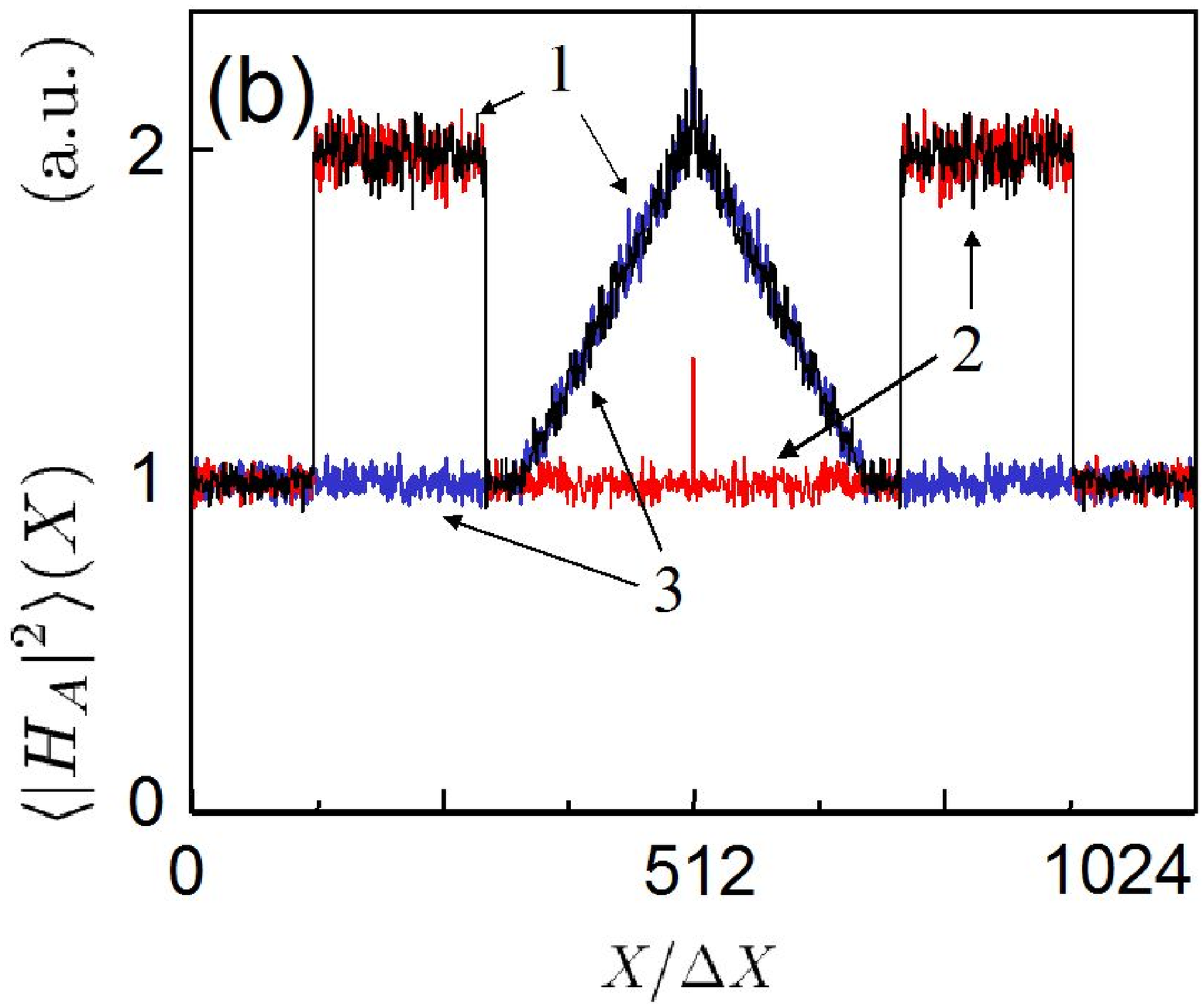}
   \caption{Curves $\langle  |H_A(X)|^2 \rangle$ calculated without (a) and with decorrelation (b) by switching on and off  the  tagged and  untagged photons signals. }\label{Fig3}
 \end{center}
\end{figure}

To confirm this analyse  of  Fig.\ref{fig2}, and to   evaluate  how   untagged photons,   shot noise and  decorrelation   affect the UOT signal,  we have calculated the curves $\langle |H_A(X)|^2 \rangle$ without and with decorrelation by switching on and off  the  tagged and  untagged photons.
To better compare results obtained without and with decorrelation,   the tagged and untagged photon energies  were measured  within a time  equal to the recording time of the sequence of $M$ frames ($2\pi M/\omega_C$) without decorrelation, and to the  frame exposure time, which is made equal to $\tau_c$, with decorrelation.
Calculations were made with  $|E_{LO}|^2= 10^4$,    $M \langle|E_{A,U}|^2 \rangle= 10000$, $M \langle|E_{A,T}|^2 \rangle= 1$  and $M=12$ without decorrelation,  and  with $|E_{LO}|^2= 10^4$,  $ \langle|E_{A,U}|^2 \rangle= 250$ and   $\langle|E_{A,T}|^2 \rangle= 1$ with  decorrelation. Note that the calculations are made  with the same  tagged photon signal with and without decorrelation (1 photo electron per pixel).

Figure \ref{Fig3} shows the curves obtained without (a)  and with (b) decorrelation.
Curves 1 (back) were obtained with  tagged and untagged  photons, curves 2 (red) with tagged  (and without untagged) photons, and curves 3 (blue) with untagged  (and without tagged) photons.
As seen, the  rectangular walls (located on the left side of Fig.\ref{Fig3}(a) and on the left and right sides of  Fig.\ref{Fig3}(b) ) correspond to the tagged photon signal.  On the hand,  the triangular walls (located  in the center of Fig.\ref{Fig3}(a) and (b) ) correspond to the untagged photons. Note that the width of  triangular walls  is twice the width of the rectangular walls, which is itself  proportional to the width of the aperture. By a proper choice of the aperture size, here and in~\cite{gross2003shot},   the rectangular and triangular walls are well separated, making possible to filter off the unwanted untagged photon signal.    Note that the effect of the untagged photons is much lower  without decorrelation. To still visualize them  in Fig.\ref{Fig3}(a), we have performed the calculation with a much larger unttagged signal without decorrelation  ($ \langle|E_{A,U}|^2 \rangle= 10^4$), than  with decorrelation  ($ \langle|E_{A,U}|^2 \rangle= 250$).

\begin{figure}
  \begin{center}
   %Requires \usepackage{graphicx}
  %\includegraphics[width=4cm]{images/fig_ima_incorr_a}
%     \includegraphics[width=4cm]{images/Fig_recons_c}
%
  \includegraphics[width=4cm]{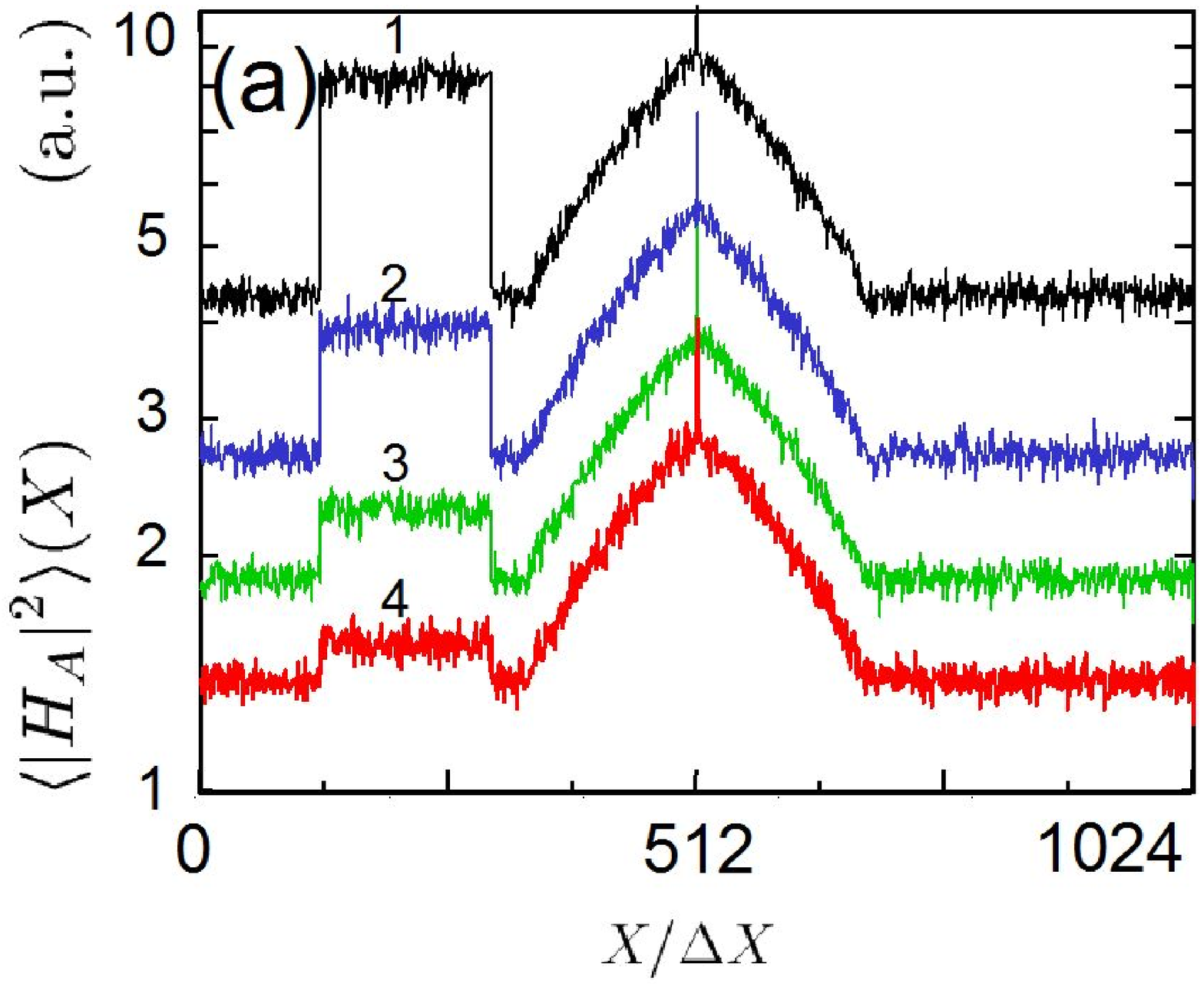}
 \includegraphics[width=4cm]{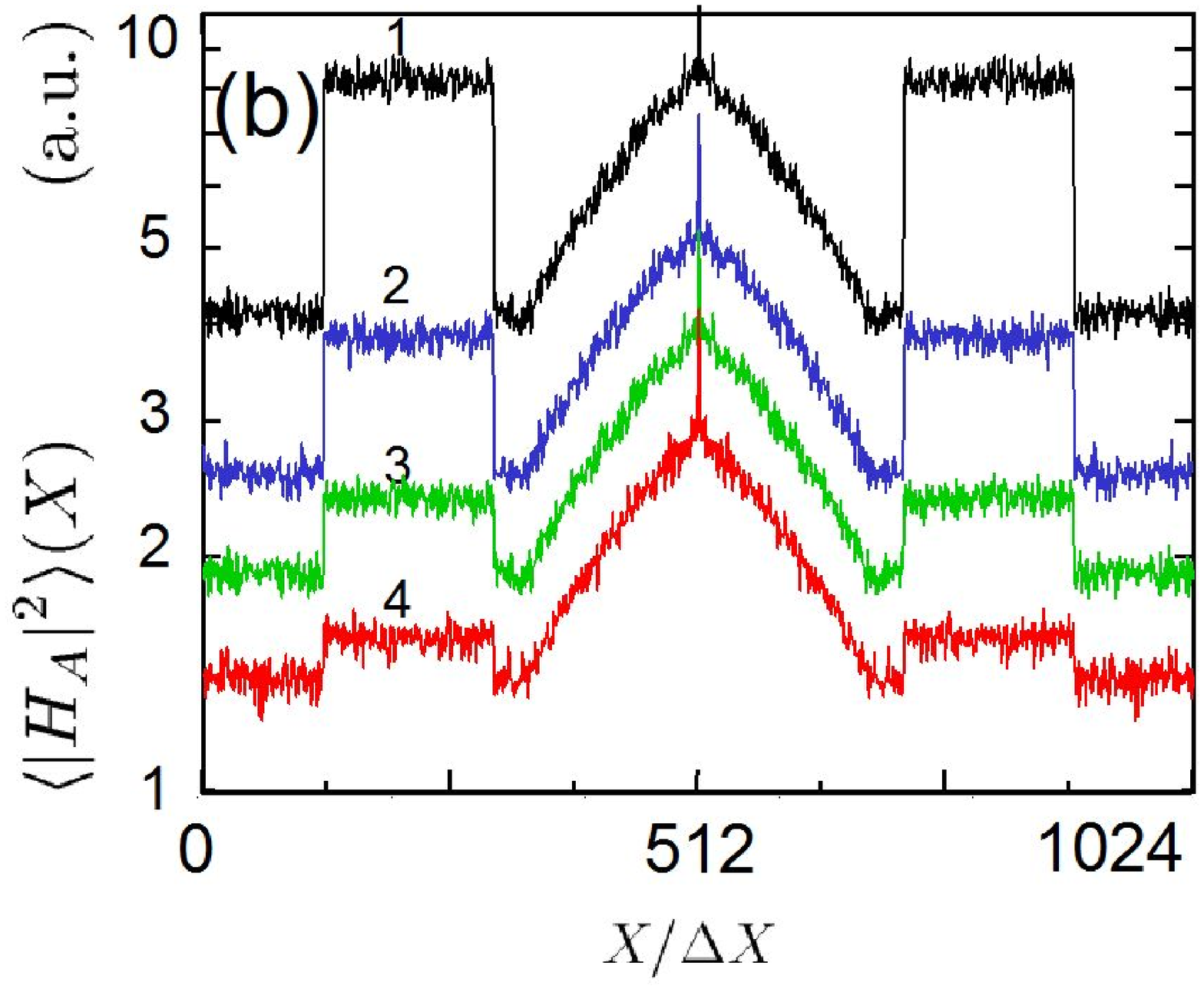}
    \caption{Curves $\langle  |H_A(X)|^2 \rangle$ obtained without (a) and with decorrelation (b). Calculation is made with $\alpha = 1$ (1), 0.5 (2), 0.25 (3) and 0.125 (4). Plots are   made in arbitrary logarithmic scale.  }\label{Fig4}
 \end{center}
\end{figure}

To evaluate tagged photon detection sensitivity limits,
we have calculated $\langle |H_A|^2\rangle (X)$ curves  by  varying the total tagged photon energy per $\tau_c$. The curves are plotted on Fig.\ref{Fig4} with decorrelation (a) and  without (b).
Calculations were made with  $|E_{LO}|^2= 10^4$,    $M \langle|E_{A,U}|^2 \rangle= 10000$, $M \langle|E_{A,T}|^2 \rangle= \alpha$  and $M=12$ without decorrelation,  and  with $|E_{LO}|^2= 10^4$,  $ \langle|E_{A,U}|^2 \rangle= 250$ and   $\langle|E_{A,T}|^2 \rangle= \alpha$ with  decorrelation, where $\alpha$  is the number of tagged photons  per pixel with $\alpha =1$, 0.5, 0.25, 0.125 and   0.0625 for curves 1 to 5.
 To better visualize them, the curves were plotted in log scale, and  the curves  were arbitrarily  shifted up or down to better separate them from each other.
The results of Fig.\ref{Fig4} show that heterodyne  holography UOT exhibits  roughly the same sensitivity for the detection of the tagged photon with and without decorrelation.   The key parameter is the tagged photon energy per pixel during the coherent measurement time  with is  equal to  $\tau_c$ with decorrelation, and to time $M 2 \pi/\omega_C$ needed to record the sequence of $M$ frames without decorrelation.  A signal versus background ratio of 1 corresponds, with and without decorrelation,  to one photo electron per pixel. By averaging over the about $10^5$ pixels of the rectangular aperture, the sensitivity limit is improved down to about $1/\sqrt{10^5} \sim 1/300$ photo electron. This result  agrees with what observed experimentally for the detection of the untagged photons \cite{gross2005heterodyne}.

In this letter, we have proposed a theoretical model to describe the detection of the tagged photons in heterodyne holography UOT. This model, which  agrees with the results of \cite{gross2003shot}, has been used to calculate how   untagged photons, speckle noise,   shot noise, decorrelation and  etendue,  affect the UOT signal. For a given coherent measurement time, which is  $M 2 \pi/\omega_C$ and  $\tau_c$ without and with decorrelation, the model yields the same  detection sensitivity,  and the same noise floor (one photon electron per pixel). Heterodyne  holography UOT is thus shot noise limited. By averaging over the $K\sim 10^5$ pixels of the  image of the aperture,   the detection sensitivity becomes $1/\sqrt{K}$ photo electron per speckle (i.e per etendue $\lambda^2$). We hope  this work will stimulate further  UOT development.

This work has been carried out thanks  to the supports of the LabEx NUMEV project (n° ANR-10-LABX-20) funded by the "Investissements d'Avenir" French Government program, managed by the French National Research Agency (ANR) and thanks to the ANR  grand  (reference n$^\circ$ANR-11-BS04-0017: ICLM).

\bibliographystyle{unsrt}

\end{document}